\documentclass[a4paper,10pt]{article}

\usepackage[utf8]{inputenc}
\usepackage{authblk}

\usepackage{graphicx}
\usepackage{url}
\usepackage{gensymb}
\usepackage{amsfonts}
\usepackage{amsmath}

\title{DF-Fit : A robust algorithm for detection of crystallographic information in Atom Probe Tomography data}
\author{Daniel Haley\footnote{Corresponding Author: daniel.haley@materials.ox.ac.uk}, Paul A. J. Bagot, Michael P. Moody}
\date{}
\affil{\small{Department of Materials, Oxford University, 16 Parks Road, Oxford, OX1 3PH, UK.}}

\begin{document}
\maketitle

\section*{Abstract}
We report on a new algorithm for detection of crystallographic information in 3D, as retained in Atom Probe Tomography (APT), with improved robustness and signal detection performance. The algorithm is underpinned by 1D distribution functions, as per existing algorithms, but eliminates an unnecessary parameter as compared to current methods.

By examining traditional distribution functions in an automated fashion in real space, rather than using Fourier transform approaches, we utilise an error metric based upon the expected value for a spatially random distribution for detecting crystallography. We show cases where the metric is able to successfully obtain orientation information, and show that it can function with high levels of additive and displacive background noise. We additionally compare this metric to Fourier transform methods, showing fewer artefacts when examining simulated datasets. An extension of the approach is used to aid the automatic detection of high-quality data regions within an entire dataset, albeit with a large increase in computational cost.

This extension is demonstrated on acquired Aluminium and Tungsten APT datasets, and shown to be able to discern regions of the data which have relatively improved spatial data quality. Finally, this program has been made available for use in other laboratories undertaking their own analyses.

\section*{Introduction}
Atom Probe Tomography (APT) is a nanoscale 3D analysis technique, which offers an otherwise unavailable combination of chemical and spatial information, across a full range of elements. The technique is applied to an increasingly wide variety of materials research areas, from metallurgical to semiconductor and even biological applications.  However, until recently, it has been difficult to perform crystallographic analyses at the atomic-scale from within atom probe datasets, although strong interest remains~\cite{Miller2012}. 

Even using state-of-the-art data analysis techniques, it can be difficult to detect the presence of crystallographic information from within an individual atom probe analysis owing to poor signal, with degradation in the form of spatial noise~\cite{Marquis2010}. The difficulty in accessing this limited information from within the atom probe data has led many authors to attempt to use correlative approaches (Electron Backscatter Diffraction (EBSD), Transmission Electron Microscopy, Transmission Kikuchi Diffraction) in order to combine crystallography information extrinsically with the atom probe's chemical fidelity~\cite{Kirchofer2014}\cite{Babinsky2014}. The extractable information using a correlative approach is invaluable, and arguably difficult to extract by other techniques~\cite{Herbig2015}.

These approaches require careful planning at the experimental design stage, and usually are limited to advanced laboratories equipped for such experimental procedures. Subsequently, post-hoc information may not be available for analyses where crystallographic information was not originally targeted. Secondly, it may be that the crystallographic information must only be extracted from small regions, making signal extraction laborious. As such, there is a clear need to expand the operating window in which crystallographic data can be extracted from atom probe datasets. This is a greater problem in reflectron equipped atom probes, owing to distortions in the flight path translating into reduced spatial resolution~\cite{Sijbrandij1996}.

To further extend the capabilities of the technique, we develop a new algorithm, DF-FIT, which utilises a real-space approach to detect the presence of crystallographic information from within a 3D APT dataset. This algorithm has fewer parameters than existing algorithms, has improved signal detection performance, and additionally is extended to provide semi-automated quantification of crystallographic information from an APT dataset.

\section*{Existing Functions}

There has been considerable literature on the topic of crystallographic determination within atom probe. Existing approaches to extract this information from APT data have utilised Hough transformations~\cite{Yao2011}, Fourier transformations~\cite{Vurpillot2001}\cite{Vurpillot2004}, radial distribution functions~\cite{Haley2009}, manual 1D and 2D distribution functions~\cite{Geiser2007}\cite{Boll2012}\cite{Moody2009}\cite{Boll2007} and semi-automated Fourier transformations~\cite{Araullo-Peters2012}. The aim of these works has often been to determine where atom probe data can provide information about any crystallinity in the dataset, such as lattice spacing, or crystal orientation.

Current distribution functions for crystallographic information extraction are computed by specifying a set of points $P$, and an analysis vector, $v$. These combine to generate a one, two or three dimensional function $F$. We restrict ourselves here to the 1D case. The function attempts to detect planes by repeatedly histogramming the difference vectors between a source point $p_i$, taken from the set of all points $P$, and a second point $p_j$, $i \ne j$, along a normal vector $v$, where the set $P_j$ is usually taken such that $||p_j - p_i|| < R$, where $R$ is the search radius parameter. Specifically, the set of vectors to atoms around the point $p_i$, denoted $\mathbb{P}_i$, is generated as follows, and the function $F(r)$, where $r$ is signed radius around the central atom $p_i$, computed from the elements of this set of vectors (pairs between atoms), $p'_k$. $\mathbf{Hist}$ is the histogram operator, here operating over the index $k$:

\begin{equation}
\begin{split}
\widetilde{\mathbb{P}_i} = \{ p_j  \in &P ~|~ ||p_j-p_i ||< R \ , i \neq j\} \\
F(r) =& \underset{k}{\mathbf{Hist}}(||p'_k|| \cdot \overrightarrow v ), ~~ p'_k \in \widetilde{\mathbb{P}_i}
\end{split}
\end{equation}

\begin{figure}
\begin{center}
 \includegraphics[width=0.9 \textwidth]{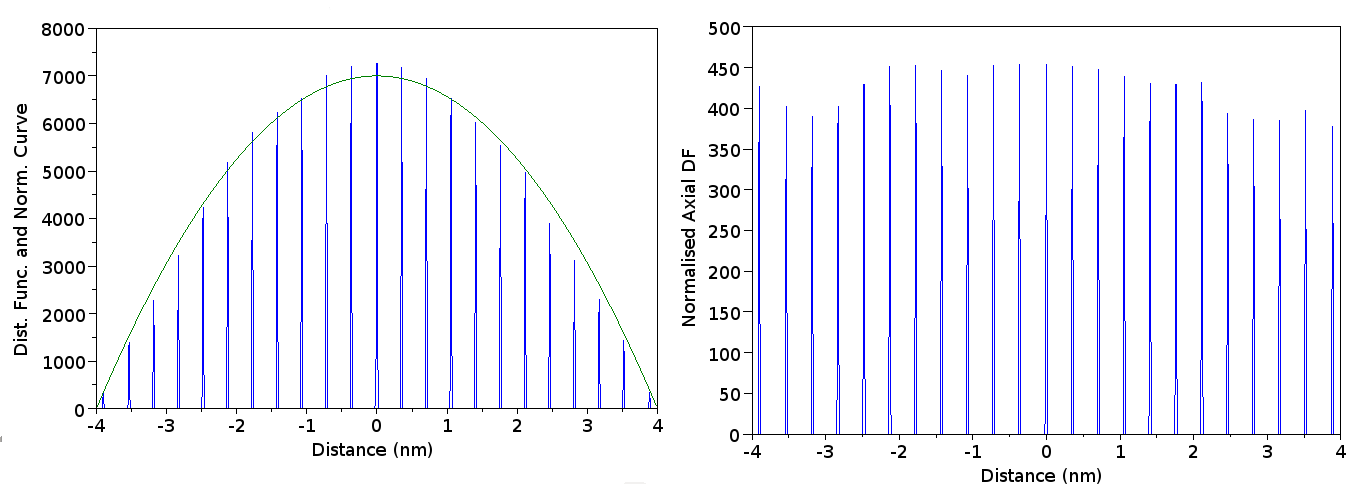}
 \caption{Unnormalised (left) and normalised (right) distribution functions for a perfect crystalline simulation. The normalisation envelope ($n \propto (R^2-r^2)$, green line), has been amplified to match the height of the distribution, for visual purposes. Fluctuation in peak height in input is due to sampling of input during function computation.} 
 \label{fig:axialDFNormalise}

\end{center}
\end{figure}

In a perfect crystalline dataset, this produces a series of Dirac delta functions when $v$ is normal to a plane, with the distance between peaks being the plane spacing, as seen in Figure~\ref{fig:axialDFNormalise}. The amplitude of these peaks is inherently given by several variables, the total number of points, $||P||$, the number of points on any given plane, and importantly the shape of the selected volume from which $p_j$ is drawn. If $P$ is a spatially random dataset (\emph{i.e.} one where points have been placed randomly in space with uniform expected density), and the points $p_j$ are drawn from a spherical volume around $p_i$, i.e.\ the aforementioned $||p_j - p_i|| < R$, then it can be shown by integration of a solid of revolution (the sphere formed by $||p_j - p_i||  < R$) that the enclosed number of points $N$ is a function of vector radius. 

As the histogram components are given by the dot product between the centre of the sphere, and the vector to the ion to match,  the final function originates from the combination of all vectors' dot products from all ions. For each ion, this is formed by making a shell of size $R$ around each ion. Thus, in a random dataset, this is equivalent to integrating a fixed volume of homogeneous density, $\rho$, which results in the following equation:

\begin{equation}
N = \rho \pi || R^2 - r^2 || \; , \;  ( ||r|| < R )
\label{eqn:axialFunction}
\end{equation}

\begin{figure}
\begin{center}
 \includegraphics[width=0.9 \textwidth]{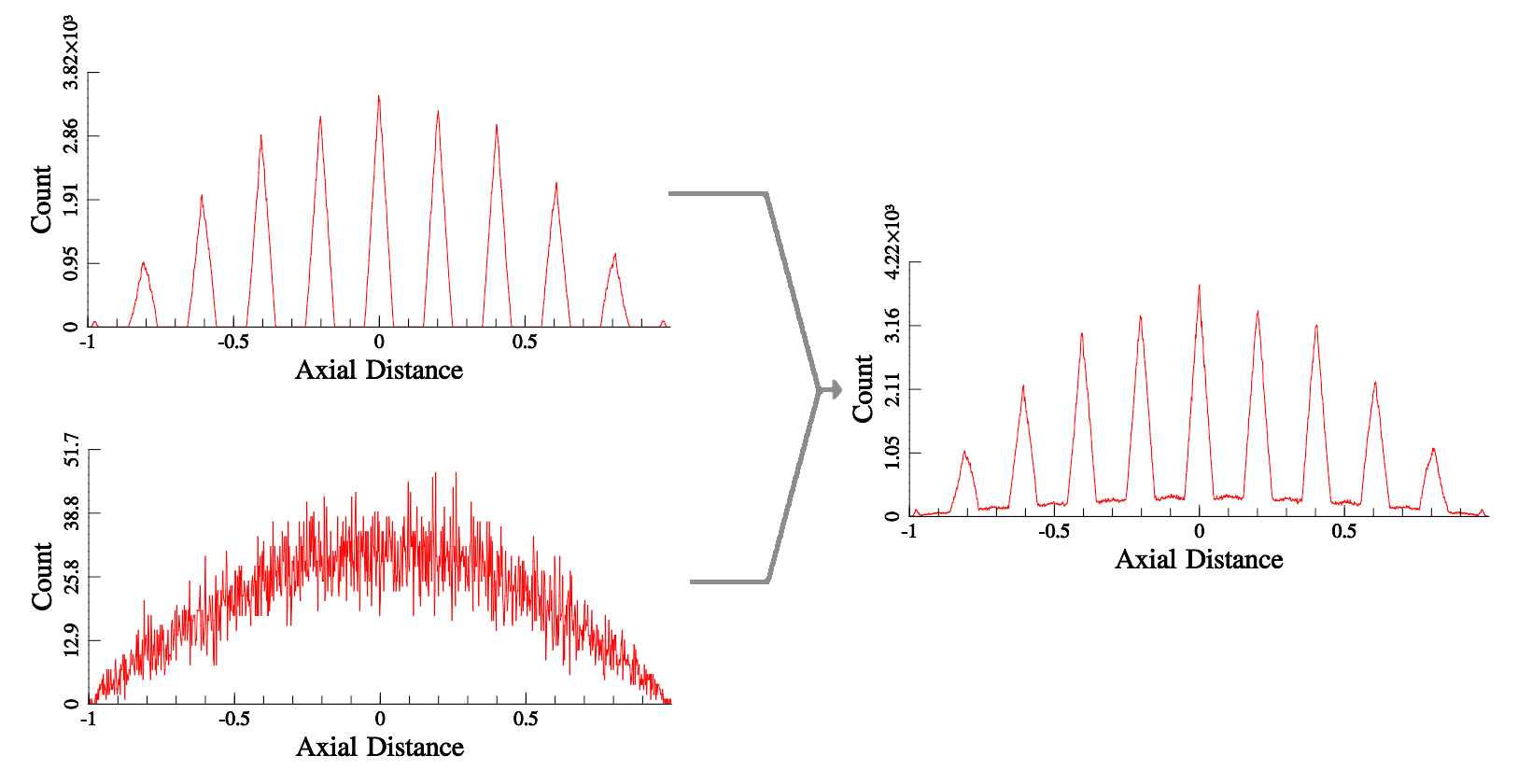}
 \caption{Axial distribution functions for simulated datasets (without
normalisation). Top left :FCC crystal with a small quantity of white displacive
noise. Bottom Left : Random additive noise only. Right: Combined additive and
displacive simulated noise.}
  \label{fig:axialNoiseEffect}

\end{center}
\end{figure}

Normalisation of this can be performed (Figure~\ref{fig:axialDFNormalise}) to allow for direct, quantitative comparisons of peak relative amplitudes between different peaks in the same plot, as a minor extension to standard algorithm discussed by Geiser~\cite{Geiser2007} and expounded by Moody~\cite{Moody2009}. Normalisation is a simple division (to account for the geometrical effects), then subtraction (to account for additive noise), to correct the distribution functions. Previous assertions by the authors that the ``background'' to the distribution function (DF) is Gaussian~\cite{Gault2010} are incorrect - the same normalisation envelope will apply to any uncorrelated noise.  Boundary effects will change the shape of the sampled volume, and thus the resultant function, but are not considered here for simplicity. It is possible to simply exclude these from the calculation of the point set $P$, thus avoiding their influence entirely.

Figure~\ref{fig:axialNoiseEffect} shows the result of this function, $F$, on Face Centred Cubic (FCC) crystals that have been modified by displacive and additive spatial noise, similar to that observed in experimental APT data, and the combination of these two results thereof. Note that as the function is simply a counting function, the effect of operation on two disjoint point sets $P$, $Q$, is simply additive. Thus ``shot'' (random additive) noise will, by definition, have a purely additive effect to the plane signal.

\section*{DF-Fit Algorithm}

Equation~\ref{eqn:axialFunction}  provides the expected value for the DF\@. As such, it is straightforwards to fit the expected value for a random distribution, with only a stochastic error from this. As a simple model, we can substitute a degree 2 polynomial for Equation~\ref{eqn:axialFunction} to fit $F_{\mathrm{fit}}(r)$ (the polynomial approximation to $F(r)$). However, if the data is not random, we cannot perform such a fit well, as there will be peaks present that are not predicted by this equation. Thus we can define a simple metric to discern the relative ``randomness'' of the observed DF, by using the residual of the fit, $E$, to quantify the error directly (Equation~\ref{eqn:errorvalue}). Therefor this error provides a direct quantification of the level of ``crystallinity'' in the set of points $P$, and in this manner we are able to detect small deviations from random data.

\begin{equation}
E^2 =  \frac{1}{N^2}\int ( F(r) - F_{\mathrm{fit}}(r) )^2
\label{eqn:errorvalue}
\end{equation}

Indeed, in a manner similar to that of Arullo-Peters~\cite{Araullo-Peters2012}, this can be computed iteratively over the spherical coordinates $(\theta, \phi)$, to produce a projection of the error over an angular range. The advantage of the representation in this work is that a direct measure of the error allows for a more robust quantification of deviations within the structure of $P$.

\section*{Performance Comparison}
\subsection{Signal quality}

First, we consider a simple simulation where a series of X-Y atom planes are generated and stacked at a fixed with apart. Atomic positions are then randomly distributed along the surface of the planes, hence there exist planes in only a single direction. This data is then rotated to an arbitrary orientation, in order to allow for retrieval by a plane detection algorithm. 

\begin{figure}
 \centering
 \includegraphics[width=0.9 \textwidth]{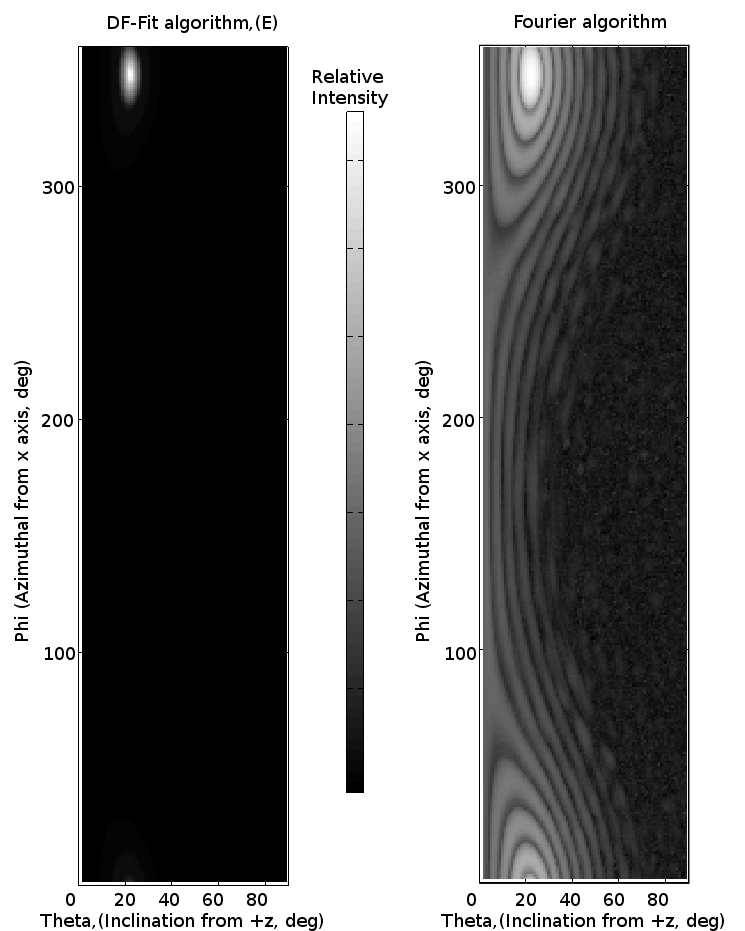}

 \caption{Comparision of DF-Fit and FFT detection algorithms on a series of stacked Z planes containing random data, rotated to an arbitrary orientation. The DF fit algorithm correctly identifies the position of the planes, as does the FFT (with damping window). However, the FFT exhibits artefacts which likely arise from edge effects in the 1D distribution function, or in 3D space. The brightness is scaled to maximum signal for each case.} 

 \label{fig:planes_df}
\end{figure}

The data can be retrieved by the FFT approach, utilising the maximum detected frequency value~\cite{Araullo-Peters2012} (10 \% natural frequency cutoff, based upon bin-size - chosen by visual inspection of FFT signal)  or by the DF-Fit algorithm, as shown by Figure~\ref{fig:planes_df}. These functions plot the estimated intensity of the signal periodicity (FFT) or deviation from randomness (DF-FIT) as a function of orientation.  

In the first case, the peak is clearly detected correctly, but not in the FFT case. Additionally ``ringing'' artefacts are visible in the FFT, but not in the DF-Fit algorithm. The application of a cosine window (harris-nuttall) used in the implementation here does not elminate these ringing artefacts. These artefacts originate from the shape of the signal itself. The signal is not truly periodic, and the plane positions of the DF can straddle the signal ``window'' which arises from the selection of the search radius, $R$, - thus a peak can be quickly repeated, or not, simply depending upon the radius that the user selected for their data analysis - this is clearly incorrect, and highlights that the assumption of periodic conditions that is inherent in the Fourier transform may not be fully appropriate. Whilst damping windows can be used to limit the transform, as shown here, this does not eliminate this effect. Indeed, such damping is unnecessary when using the DF-Fit approach, and is an additional concern in the FFT approach. As such, the DF-Fit approach removes a parameter - the cut-off frequency used to suppress ``noise'' (low-frequency effects from the DF's overall shape) and eliminates the need for windowing - a clear advantage. 

Secondly, we consider the case where spatially random noise and FCC data are simulated concomitantly. In APT datasets, there is positional noise (errors in atomic positions) due to e.g.\ abberations in trajectories and errors in positinoning during reconstruction, but furthermore there can be additional ``ions'' erroneously added to the dataset, due to factors such as signal-to-noise concerns, background noise, ``overlaps'' in mass spectra and mass-peak tails.

The noise is simulated as additive shot noise and as random offsets in the atomic positions. The algorithm implemented here and the FFT algorithm are compared when considering an anisotropically blurred FCC crystal (Gaussian noise, applied to the position of each atom) with a significant number of noise events adjacent ($\sim$10 noise atoms per crystal atom).

The algorithm was run with the same parameters on the input, using a 10~\% frequency cutoff in the FFT case, with no window function. Additionally, a simple peak detection algorithm is used to label the most prominent maximum peaks. Specifically, the Octave \emph{immaximas} routine was  used~\cite{OctaveForge16}, which searches local spatial maxima within a square region using a parabolic fit to obtain sub-resolution accuracy, with thresholding to suppress non-prominent peaks. The DF-Fit algorithm shows clearly detectable peaks, whereas only the main peak is distinguishable from surrounding noise in the FFT case. 

\begin{figure}
 \centering
 \includegraphics[width=0.8 \textwidth]{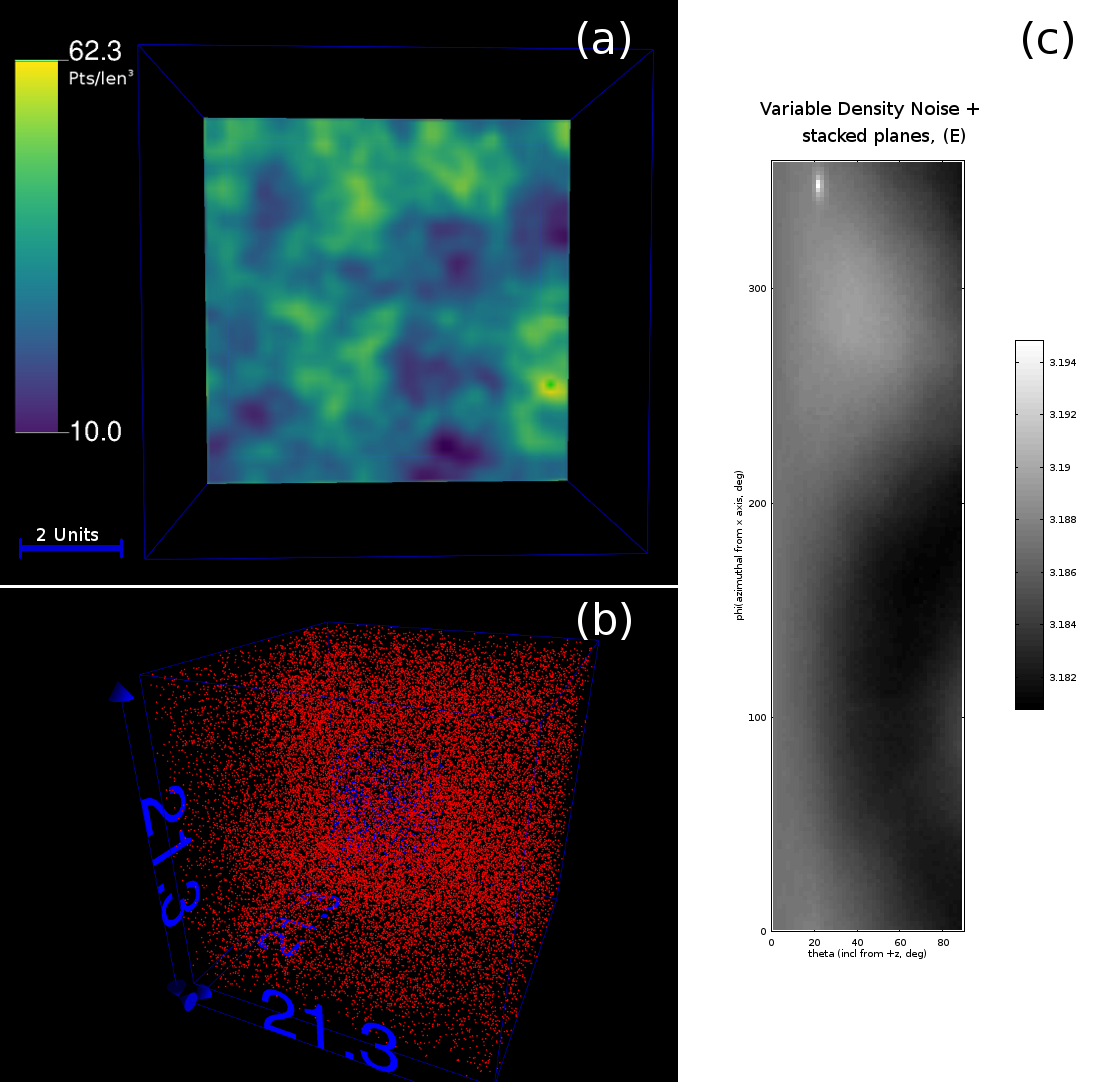}
 \caption{Usage of DF-FIT with stacked Z-Planes from Figure~\ref{fig:planes_df}, with additive variable density (``Perlin'') noise. (a) shows the density of noise atoms in the dataset, (b) showsn the Z-planes embedded within the noisy data, and (c) shows the DF-FIT algorithm correctly extracting the data. In this example there are 2 million noise points, and 650,000 ``crystal'' points.} 

 \label{fig:perlinNoise}
\end{figure}

Lastly, we examine the case whereby there is a varying, isotropic, noise density within the dataset. This is modelled as so-called ``Perlin'' noise, as shown in Figure~\ref{fig:perlinNoise}(a), in a 21 unit length box, of 2 million points. Superimposed into this box is the data from Figure~\ref{fig:planes_df}, which consists of 62500 points - this combined dataset is shown in Figure~\ref{fig:perlinNoise}(b). Thus there is a ratio of $\sim$0.031 `signal' to `noise' atoms within the dataset. As the noise with varying density is isotropic, it does not interfere with the detection of the orientation of the planes in the dataset, as shown in Figure~\ref{fig:perlinNoise}(c). Thus it is still possible to detect the orientation of crystallographic planes, even when there are variations in density. This result is largely because firstly, the density variations are not large in the local search area, and secondly as these variations are isotropic.

\subsection*{Computational speed}
At the conditions shown in Figure~\ref{fig:planes_df} (1000 sample ions for DF generation, 1 degree angular resolution, $\sim$87,000 input points), the total run-time for the program is approximately 130 seconds (Intel i5-5470, 3.4~GHz), where run-time changes with the square of angular resolution, cube of radius, and is linear in input sampling, and $n \log (n)$ in total points (elsewhere it is quoted as linear~\cite{Marquis2010}, but this does not account for a requisite $\log(n)$ NN search time). The quoted run time is for a parallelised program using a KD-tree~\cite{Moore1991} based neighbour search. For most inputs the algorithm is limited in speed by the neighbour search, not by the signal analysis technique - as such the FFT and DF-Fit operate at similar speeds. As a performance improvement when computing the angular map, the probe ion's neighbours are computed once each for all vectors in the angular range ($\theta$, $\phi$) and thus  the entire angular map for that ion is accumulated from a single neighbour search.

\begin{figure}
 \centering
 \includegraphics[width=0.9 \textwidth]{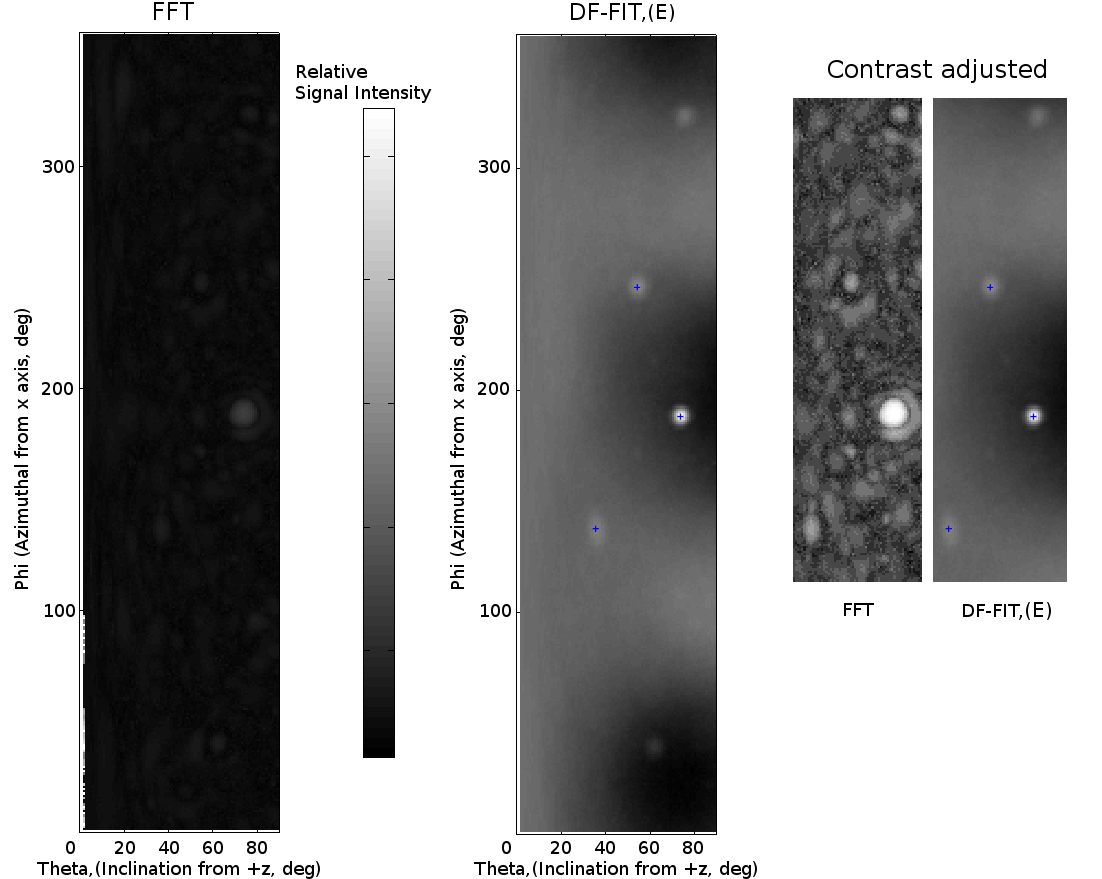}
 
 \caption{Comparison of FFT and DF-Fit algorithms on FCC crystal with anisotropic and shot noise ($a=0.405, x=0.02,y=0.05,z=0.05$, 7000 ions with 80,000 ions shot noise (non-overlapped)). Image brightness (left pair) set to maximum pixel. Brightness on right pair set to match contrast in each case. Markers  (+) show automatic peak detection in each image using Octave's \emph{immaximas} routine  (same parameters) - no peaks are automatically detected in FFT case.}
 \label{fig:df_fit_compare}
\end{figure}

\section*{Applications}

As the signal performance of the DF-Fit algorithm is improved over that of the FFT, it is possible to utilise this enhanced detectability to allow for more complex analyses. 

An Al wire sample was prepared using standard electropolishing methods (2~\% Perchloric in Butoxyethanol fine), and then analysed on a LEAP3000-HR system, run in voltage pulsing mode at 35~K (set point), 15~\% pulse fraction and 200~kHz pulse repetition frequency, with a total dataset size of 50~M ions. The dataset was reconstructed using default parameters from the IVAS software package (3.6.12). 

\begin{figure}
 \centering
 \includegraphics[width=0.9 \textwidth]{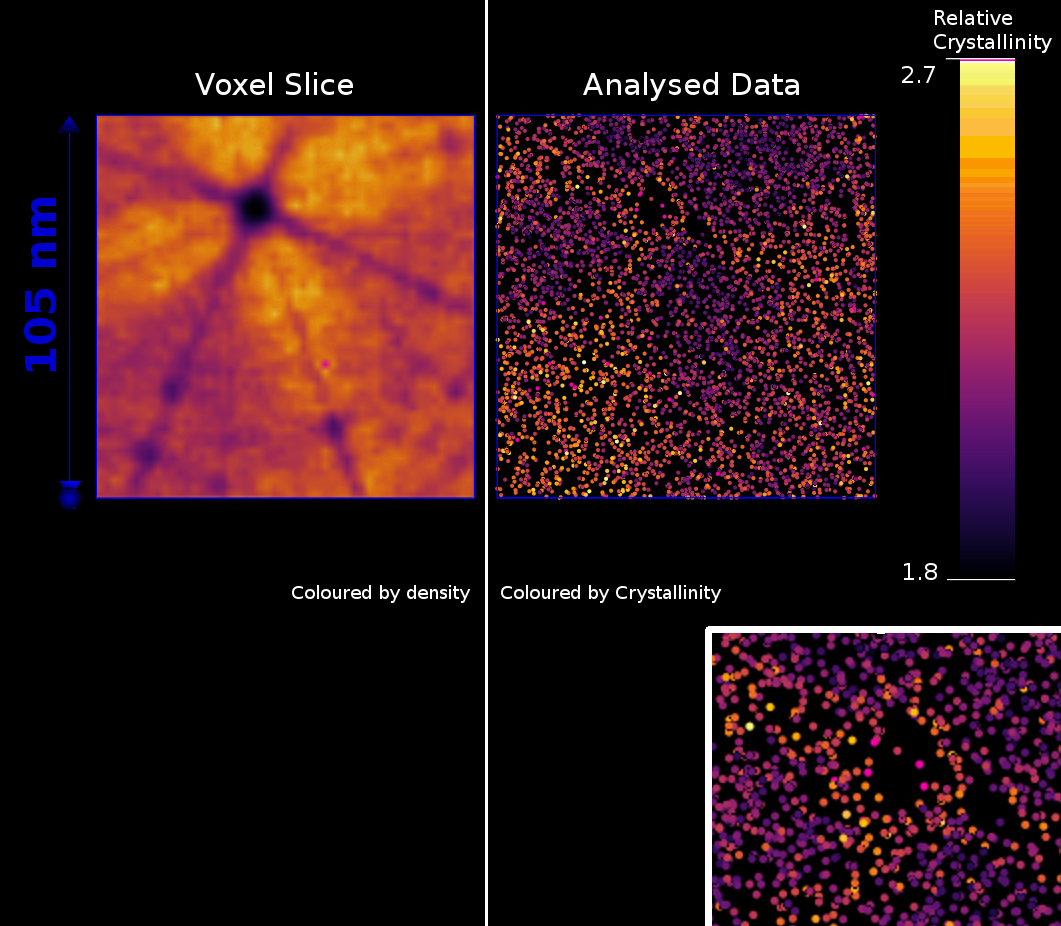}
  \caption{Relative ``crystallinity'' measure, as computed using the DF-fit algorithm, repeated over small sampling volumes. Original APT data acquired from an aluminium wire at 30~K. Inset shows close-up of pole region, slice thickness is 30~nm.}
 \label{fig:crystallinity}
\end{figure}

A simple approach was utilised to incorporate the DF-Fit algorithm into a larger algorithm to determine regions of the data that have a strong crystallographic signal. Random ions were selected from the dataset to create a ``probe volume'', at each of these points a spherical volume was extracted, and this was analysed using the DF-Fit algorithm. From each probe volume, a 2D map is generated (similar to Figure~\ref{fig:df_fit_compare}, where the brightness of the pixels in the map indicates a high degree of data quality. For simplicity, the most prominent pixel was selected as the representative value for crystallographic signal quality. This was repeated to generate points in 3D space, to build a 3D image of the quality of the crystallographic dataset (Figure~\ref{fig:crystallinity}). It is not required to sample this using the original points -- a regular 3D grid could be used -- however using the dataset itself as the support for this calculation provides the advantage that it naturally adapts to the datasets own density.

As is to be expected, the area just outside the major pole shows a strong signal - highlighting the presence of crystallographic information. This is true despite the limited peak detection implemented in this work (maximum pixel only). This could be further extended to provide a 3D vector map, showing the orientation of the strongest pole, as this information has already been computed. This allows for analysts to automatically determine the locations within their datasets that yield the best crystallographic information, removing the need for previously tedious manual searches. 

\begin{figure}
 \centering
 \includegraphics[width=0.9 \textwidth]{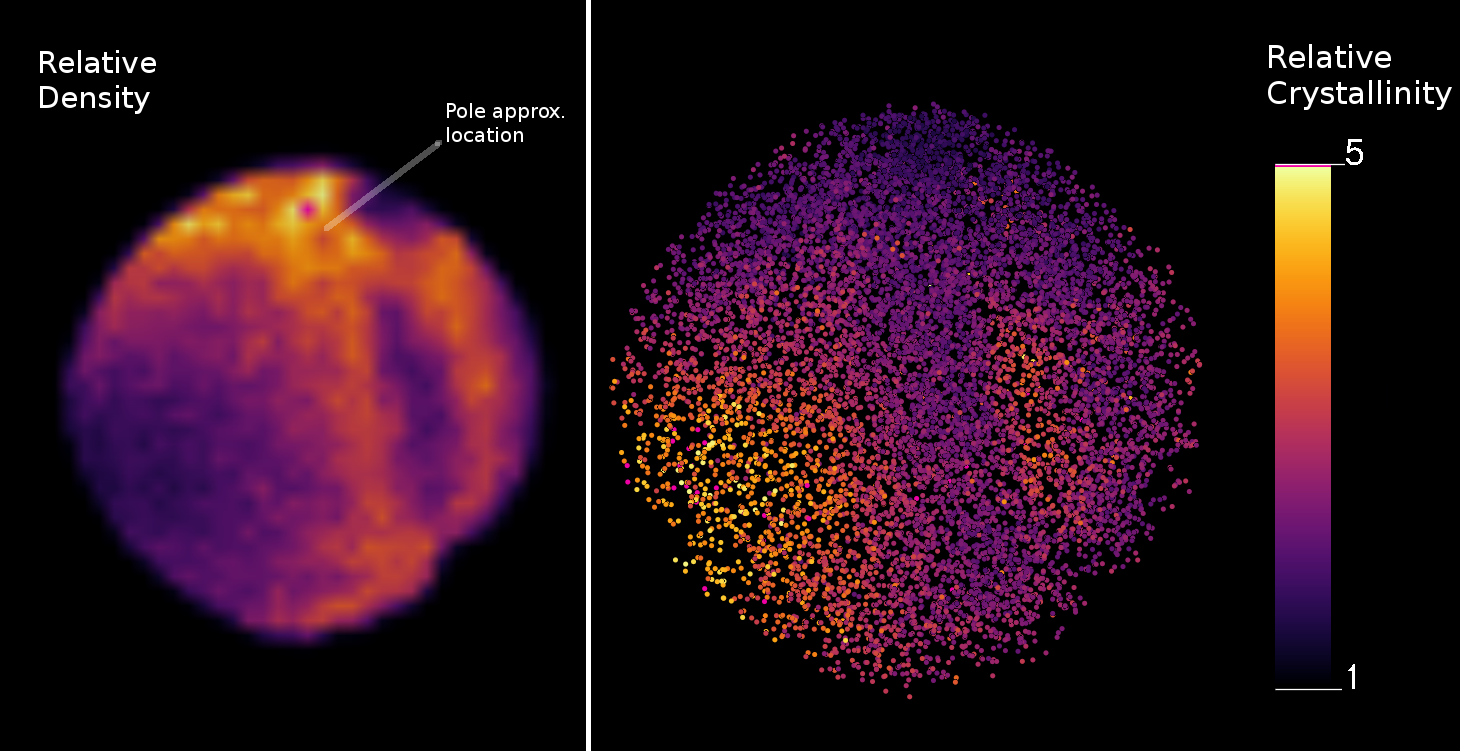}
 \caption{Crystallinity analysis of W-Irradiated W (6~Dpa, 500 \degree C). Unexpectedly, the pole locations are not the ideal locations for crystalline information. Note that the surface has been cropped in the out-of-page direction, to remove artefacts from boundary effects.}
 \label{fig:irradiatedw}
\end{figure}

As a second example, an automated analysis of the crystallinity metric (Equation~\ref{eqn:errorvalue}) was performed on a tungsten sample that had been W self-irradiated in needle form, to 6~Dpa at 500 \degree C~\cite{Dagan2015}. A comparison of the relative spatial density (left, voxel slice) and crystallinity metrics (right) are given in Figure~\ref{fig:irradiatedw}. Surprisingly, the regions with the highest ``crystallinity'', unlike in the Al sample, are not co-located to pole or zone-lines, which may be unexpected when searching for crystalline information within an APT dataset.
  
\section*{Limitations and Future work}

The approach is, however, not without some limitations. The primary limitation is, as with the FFT approach, a moderate run time. Drastic reductions in the input point count, search radius or the angular sampling range parameters need to be made to reduce analysis times to tractable levels. However, cluster computation approaches may be of benefit here, as the algorithm is simple to distribute. This is most important for the more computationally intensive ``crystallinity'' quantification algorithm.

From a numerical perspective, a concerning limitation is the effect of boundaries, or extreme changes in density. These can provide a solution that is poorly approximated by Equation~\ref{eqn:axialFunction}. The result here is that these contributions to the angular map will be distorted by a non-uniform background, as seen in the slowly varying background in the DF-Fit output in Figure~\ref{fig:df_fit_compare}. If this is more severe, this makes peaks difficult to identify, and can reduce the algorithm's numerical performance.  It is, however, unlikely that these will produce false-positive peaks in the angular map. This problem can be removed at the cost of rejecting surface data, and constraining the calculation only to atoms that are sufficiently far from the dataset edge.

The most important difference between the DF-Fit and the FFT approaches is that the DF-Fit approach does not yield any spacing information - only whether planes are either absent or present. Future work may wish to examine autocorrelation or using FFT approaches after DF-Fit analysis to determine the spacing information once the directions are identified. 

Given experiments with sufficient crystallographic quality, a useful window of opportunity is now available. The output generated from this program is of similar form to that utilised in EBSD analyses~\cite{Pinard2011}. If successfully merged with an orientation to crystal symmetry analysis tool, this would allow for fully automatic EBSD-style phase identification in an APT dataset.

\section*{Availability}

The program is available online as source code at \url{http://apttools.sourceforge.net}, or at time of print (direct): \url{https://sourceforge.net/p/apttools/extras/code/ci/default/tree/}. Data is available via the Oxford Research Archive at \url{https://deposit.ora.ox.ac.uk/datasets/uuid:a62d069f-443b-4bd7-8433-5602f8edb7fe/file/content1544bz65p}.

\section*{Conclusions}

We have developed a new method for the determination of plane directions in APT datasets. Here we have shown that this can enhance the detection of crystallographic information, above and beyond that of existing algorithms. Simultaneously we have eliminated an input parameter from the process, as the new method is more direct and robust than FFT approaches. It is anticipated that such real-space approaches may become the primary method for determining the quality of crystallographic information present in APT datasets.

\section*{Acknowledgements}

We wish to acknowledge the EPSRC Hems project  EP/L014742/1. Dr. Dagan is thanked for providing the W-Irradiated dataset.

\bibliography{dffit}
\bibliographystyle{plain}

\end{document}